\documentclass[english,runningheads,a4paper]{llncs}[2018/03/10]

\usepackage[ngerman,main=english]{babel}
\addto\extrasenglish{\languageshorthands{ngerman}\useshorthands{"}}
\usepackage{regexpatch}
\makeatletter
\edef\switcht@albion{%
  \relax\unexpanded\expandafter{\switcht@albion}%
}
\xpatchcmd*{\switcht@albion}{ \def}{\def}{}{}
\xpatchcmd{\switcht@albion}{\relax}{}{}{}
\edef\switcht@deutsch{%
  \relax\unexpanded\expandafter{\switcht@deutsch}%
}
\xpatchcmd*{\switcht@deutsch}{ \def}{\def}{}{}
\xpatchcmd{\switcht@deutsch}{\relax}{}{}{}
\edef\switcht@francais{%
  \relax\unexpanded\expandafter{\switcht@francais}%
}
\xpatchcmd*{\switcht@francais}{ \def}{\def}{}{}
\xpatchcmd{\switcht@francais}{\relax}{}{}{}
\makeatother

\usepackage{ifluatex}
\ifluatex
  \usepackage{fontspec}
  \usepackage[english]{selnolig}
\fi

\iftrue 
  \ifluatex
  \else
    \usepackage[%
      rm={oldstyle=false,proportional=true},%
      sf={oldstyle=false,proportional=true},%
      tt={oldstyle=false,proportional=true,variable=false},%
      qt=false%
    ]{cfr-lm}
  \fi
\else
  \ifluatex
  \else
    \usepackage{newtxtext}
    \usepackage{newtxmath}
    \usepackage[zerostyle=b,scaled=.9]{newtxtt}
  \fi
\fi

\ifluatex
\else
  \usepackage[T1]{fontenc}
  \usepackage[utf8]{inputenc} 
\fi

\usepackage{graphicx}

\usepackage{upquote}

\usepackage{booktabs}

\usepackage{paralist}

\usepackage{csquotes}

\usepackage{textcmds}

\usepackage{multirow}

\RequirePackage[%
  babel,%
  final,%
  expansion=alltext,%
  protrusion=alltext-nott]{microtype}%
\DisableLigatures{encoding = T1, family = tt* }

\usepackage{url}
\makeatletter
\g@addto@macro{\UrlBreaks}{\UrlOrds}
\makeatother

\usepackage{xcolor}

\usepackage{listings}
\renewcommand{\lstlistingname}{List.}
\usepackage{chngcntr}
\AtBeginDocument{\counterwithout{lstlisting}{section}}

\usepackage{pdfcomment}
\usepackage[%
  square,        
  comma,         
  numbers,       
  sort&compress, 
]{natbib}

\ifluatex
\else
  \SetExpansion
  [ context = sloppy,
    stretch = 30,
    shrink = 60,
    step = 5 ]
  { encoding = {OT1,T1,TS1} }
  { }
\fi

\usepackage{stfloats}
\fnbelowfloat

\usepackage{hyperref}
\hypersetup{hidelinks,
  colorlinks=true,
  allcolors=black,
  pdfstartview=Fit,
  breaklinks=true}
\usepackage[all]{hypcap}

\usepackage[group-four-digits,per-mode=fraction]{siunitx}

\usepackage[capitalise,nameinlink]{cleveref}
\usepackage{iflang}
\IfLanguageName{ngerman}{
  \crefname{table}{Tab.}{Tab.}
  \Crefname{table}{Tabelle}{Tabellen}
  \crefname{figure}{\figurename}{\figurename}
  \Crefname{figure}{Abbildungen}{Abbildungen}
  \crefname{equation}{Gleichung}{Gleichungen}
  \Crefname{equation}{Gleichung}{Gleichungen}
  \crefname{listing}{\lstlistingname}{\lstlistingname}
  \Crefname{listing}{Listing}{Listings}
  \crefname{section}{Abschnitt}{Abschnitte}
  \Crefname{section}{Abschnitt}{Abschnitte}
  \crefname{paragraph}{Abschnitt}{Abschnitte}
  \Crefname{paragraph}{Abschnitt}{Abschnitte}
  \crefname{subparagraph}{Abschnitt}{Abschnitte}
  \Crefname{subparagraph}{Abschnitt}{Abschnitte}
}{
  \crefname{section}{Sect.}{Sect.}
  \Crefname{section}{Section}{Sections}
  \crefname{listing}{\lstlistingname}{\lstlistingname}
  \Crefname{listing}{Listing}{Listings}
}

\usepackage{xspace}

\DeclareFontFamily{U}{MnSymbolC}{}
\DeclareSymbolFont{MnSyC}{U}{MnSymbolC}{m}{n}
\DeclareFontShape{U}{MnSymbolC}{m}{n}{
  <-6>    MnSymbolC5
  <6-7>   MnSymbolC6
  <7-8>   MnSymbolC7
  <8-9>   MnSymbolC8
  <9-10>  MnSymbolC9
  <10-12> MnSymbolC10
  <12->   MnSymbolC12%
}{}
\DeclareMathSymbol{\powerset}{\mathord}{MnSyC}{180}

\ifluatex
\else
  \input glyphtounicode
\fi

\usepackage[math]{blindtext}
\usepackage{mwe}
\makeindex

\begin{document}

\title{Automated Left Ventricle Dimension Measurement in 2D Cardiac Ultrasound via an Anatomically Meaningful CNN Approach}
\titlerunning{Automated left ventricle dimension measurement in 2D cardiac ultrasound}


\author{%
     Andrew Gilbert \inst{1}\textsuperscript{,}  \inst{2}\and
     Marit Holden\inst{3} \and
     Line Eikvil \inst{3} \and
     Svein Arne Aase \inst{1} \and
     Eigil Samset \inst{1}\textsuperscript{,} \inst{2} \and
	Kristin McLeod \inst{1}
 }

%
\authorrunning{A. Gilbert et al.}
  \institute{
      GE Vingmed Ultrasound, GE Healthcare \and 
      Department of Informatics, University of Oslo \and
     Norwegian Computing Center
}

\maketitle

\begin{abstract}
Two-dimensional echocardiography (2DE) measurements of left ventricle (LV) dimensions are highly significant markers of several cardiovascular diseases. These measurements are often used in clinical care despite suffering from large variability between observers. This variability is due to the challenging nature of accurately finding the correct temporal and spatial location of measurement endpoints in ultrasound images. These images often contain fuzzy boundaries and varying reflection patterns between frames. In this work, we present a convolutional neural network (CNN) based approach to automate 2DE LV measurements. Treating the problem as a landmark detection problem, we propose a modified U-Net CNN architecture to generate heatmaps of likely coordinate locations. To improve the network performance we use anatomically meaningful heatmaps as labels and train with a multi-component loss function. Our network achieves 13.4\%, 6\%, and 10.8\% mean percent error on intraventricular septum (IVS), LV internal dimension (LVID), and LV posterior wall (LVPW) measurements respectively. The design outperforms other networks and matches or approaches intra-analyser expert error.
\end{abstract}


\section{Introduction}\label{sec:intro}
Ultrasound imaging is the primary imaging modality used to assess cardiac morphology and function. Compared to other imaging modalities (e.g. MRI and CT), ultrasound imaging has a lower cost, is easier to perform, and, unlike CT, does not produce ionizing radiation. This makes it ideally suited for rapid diagnostic use for patients with cardiovascular disease. A diagnosis is made by acquiring a set of images from different views of the heart and extracting measurements of heart function from those images. Some of the most frequent measurements in patient care settings are measurements of the left ventricle (LV) from the parasternal long-axis view. The typical set of measurements consists of the length of the intraventricular septum (IVS), left ventricular internal dimension (LVID), and left ventricular posterior wall (LVPW) at both the end-diastole (ED) and end-systole (ES) phases of the cardiac cycle. Several examples of these measurements are shown in Fig. \ref{fig:results}. Because LV dimension measurements are performed frequently, automated measurement tools could provide tremendous time savings for clinical use. 

	Despite its widespread use, there is a high variability in LV dimension measurements due to variations in training and the difficulty of precisely detecting relevant structures. The 2010 HUNT study \cite{Thorstensen} measured inter-analyser (difference between experts reading the same exam) and intra-analyser (difference between the same expert reading the same exam several weeks apart) for several standard echocardiographic measurements. The intra-analyser mean percent error (MPE) for IVS, LVID, and LVPW measurements was 10\%, 4\%, and 10\% respectively and inter-analyser results were similar. For IVS and LVPW measurements this corresponds to about half of the standard deviation of normal ranges \cite{Kou2014} so a patient on the borderline could easily be put in a different diagnostic group. The high variability highlights the difficulty of the task at hand, but effective automation is one promising approach to reduce this variability and implement a more reproducible diagnostic pipeline.	
 
	Previous work on 2D ultrasound measurements has focused on individual measurements. Snare et al. used deformable models with Kalman filtering to outline the septum shape \cite{Snare2012}, achieving bias and standard deviation of 0.14 ± 1.36 mm for automated IVS measurements compared to manual measurements. Baracho et al. used perceptron style neural networks and filtering to generate a septum segmentation \cite{Baracho2016}. They achieved results of 0.5477mm ± 0.5277mm for IVS measurements but failed to validate directly against measurements from an expert cardiologist. Finally, Sofka et al. developed an automated method for detecting LVID measurements using convolutional neural networks (CNNs) \cite{Sofka2017}. Sofka et al. introduce a center of mass layer to regress keypoint locations and achieved a 50th percentile error of 4.9\% and a 95th percentile error of 18.3\%.We extend the work of Sofka et al. by targeting the IVS and LVPW measurements in addition to LVID. Including more measurements increases the difficulty of the task because the network should not only achieve high accuracy on all measurements but also find measurement vectors that have a logical relationship to each other (i.e. all measurement vectors should be parallel to follow clinical guidelines). Additionally, the upper IVS and lower LVPW endpoints do not fall at distinct gradient boundaries within the image making them more difficult to find, even for an expert. 

	As with Sofka et al., we frame the task as a landmark detection problem, where the goal is to identify 6 key points (the 2 endpoints of IVS, LVID, and LVPW measurements) from an input image. A landmark based approach was chosen to increase user-interpretability and allow editing of the found points by users in a clinical workflow. Many architecture variants have been applied in previous work on landmark detection problems, but the most common approach is to generate a heatmap of likely locations for each key point of interest \cite{Tompson,B2016,Nibali2018}. The heatmap is directly compared to a reference heatmap generated from the key point’s known location, or the coordinates of the key points are regressed from the heatmap and compared to known coordinates.  

  We propose several modifications to the general landmark detection strategy above because, in contrast to facial recognition, there is no defined local appearance of these landmarks. Instead, their location is determined from local appearance and global structural information. For example, while the septum typically extends through a large part of the image, ASE guidelines recommend measuring at the level of the mitral valve leaflets \cite{Lang2015} which means an algorithm needs to be aware of structural information to find the correct IVS endpoints. 

The novelty of our approach lies in it's ability to handle these challenges and achieve high accuracy. First, we generate anatomically meaningful ground truth heatmaps which follow the expected spatial distribution of the point. Second, we propose the integration of coordinate convolution layers \cite{Liu} within feature detection networks for medical imaging. Third, we optimize network performance using a multi-component loss function which provides feedback to the network in multiple components including measurement endpoint coordinate locations, angle of measurement, and measurement distances. Including all these terms allows us to optimize for both measurement accuracy and a logical relationship between measurement vectors. Finally, we evaluate several different architectures within the constraints of our first two contributions to show the optimal architecture for the given task. 

\section{Methods}
\label{sec:methods}

\subsection{Network}\label{sec:network}

 The input to the proposed network is a single 2D frame. The accurate detection of ES and ED frames from a full cardiac loop is left for future work. The image is first passed through a CoordConv layer, which adds pixel-wise spatial location information to allow CNNs to more easily find objects \cite{Liu}. The core of our approach is a U-Net \cite{Ronneberger}. A U-Net is a CNN with a sequence of down and up sampling paths with skip connections concatenating each down-sampling output to the corresponding up-sampling level. In each successive down-sampling layer, the number of filters doubles and the spatial resolution in each dimension is cut in half, while the reverse is true in up-sampling. We make several modifications in our implementation. The number of down-sampling levels and the number of filters are parameterized to tune the network. Padding is added on all layers to ensure output heatmap resolution matches the input. Batch normalization and spatial dropout layers are included between convolutional blocks for regularization, avoiding standard dropout since neighboring pixels are strongly correlated \cite{Tompson}. Each convolutional layer uses a kernel size of 3x3. 

\begin{figure}
  \centering
  \includegraphics[width=\textwidth]{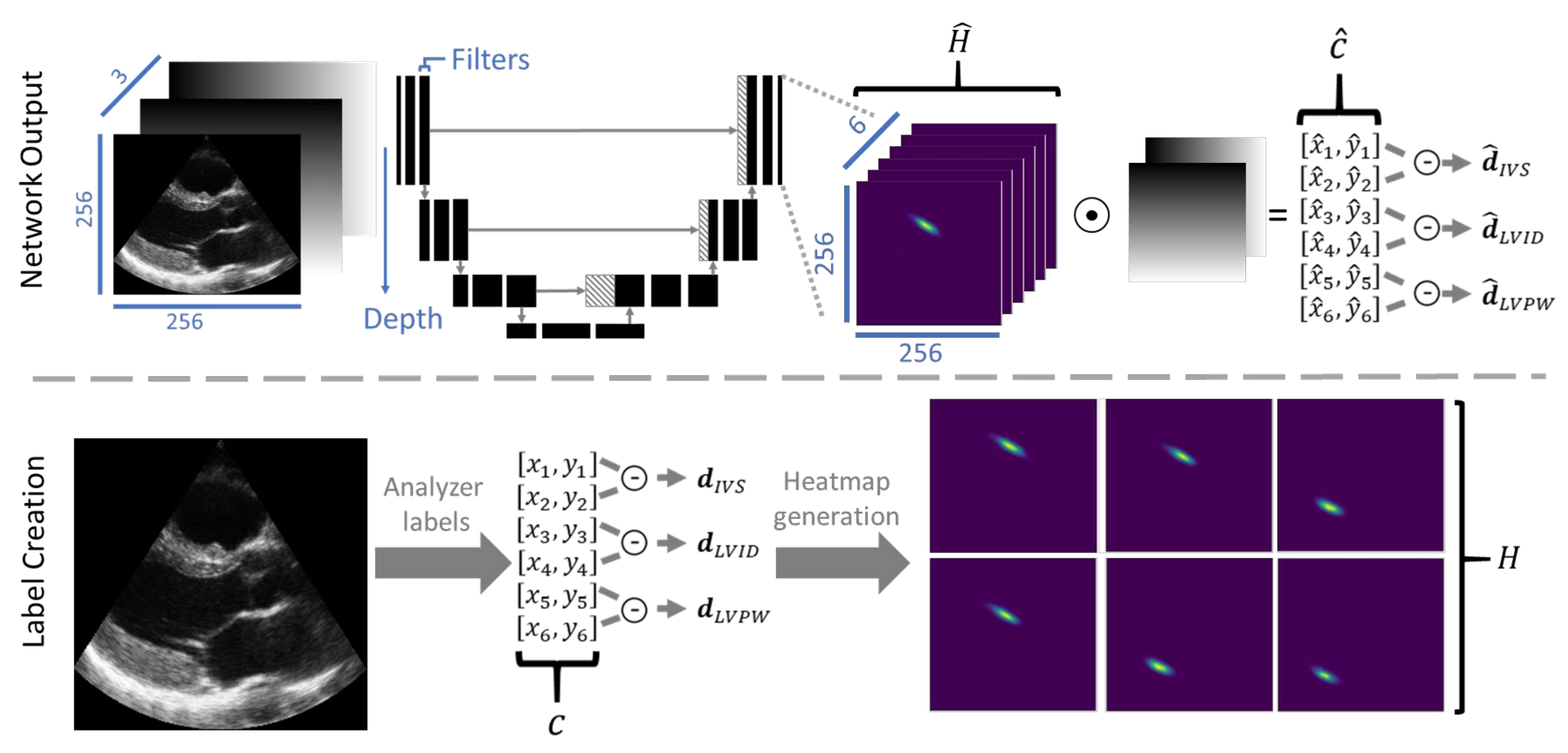}
  \caption{Network architecture. The input image (256x256) is appended with $x$ and $y$ coordinate channels to create a 3 channel image and passed through a U-Net-based architecture. The output contains 6 heatmaps ($\hat{H}$), one for each detected landmark. The center of mass of each heatmap is extracted as the found coordinates ($\hat{c}$), and vectors for each measurement are obtained ($\vec{\hat{d}}$). Label distances ($\vec{d}$) and heatmaps ($H$) are generated from labeled endpoints ($c$) to compare to the network output. 
}
  \label{fig:design}
\end{figure}

Our output is the same size as the original image but contains 6-channels, with each channel representing a heatmap corresponding to one landmark. Although the top and bottom endpoints of LVID typically match the bottom of IVS and top of LVPW respectively, they can be different for some pathologies which is the reason they are independent points in our framework. Each channel is normalized to be a probability map and passed through a differential spatial-numerical transform block \cite{Nibali2018} to calculate the center of mass in x and y: the endpoints of the three measurement vectors. From the coordinate endpoint locations, we calculate the final distance measurements. The network architecture is shown in Fig. \ref{fig:design}. 

\subsection{Loss Function}\label{sec:loss}
Our labels are the coordinate locations of all caliper endpoints. We extrapolated these to match the network output including heatmaps of coordinate locations, and distances between coordinate pairs. For the label heatmaps, a 2D gaussian is centered at the location of the labeled coordinate. The gaussian is elongated in one dimension with a ratio of 20 to 1 between the variances of the long and short axes and rotated such that the long axis was orthogonal to the direction of measurement (see $H$ in Fig. \ref{fig:design} for example). This both followed the expected spatial distribution of the points and gave the network feedback that a miss orthogonal to the direction of measurement was more acceptable than one parallel to the measurement, which would substantially affect measurement results. The variance of the gaussian in the long axis is 14 pixels (or ~5\% of the image size). 

L2 loss is used for the six coordinate locations and three distance measurements, although the distance loss was divided by the relative actual distance ($\vec{d}$) to equally weight each measurement. The heatmap loss is the root mean squared error (RMSE) between the generated and output heatmaps, following Newell et al. \cite{B2016}. The heatmap loss helps the network converge to a reasonable result quickly, because feedback is provided to the network at every pixel in the output, rather than just a single metric fed back to all pixels such as with the distance or coordinate measures. The difference in the relative angles of the measurement vectors is also included in the loss function as the cosine similarity between the two vector sets. Including the angle loss is critical because even if the network can correctly find point delineations across the relevant structure (e.g. septum), if the measurement vector is not orthogonal to that structure then the measurement will be overestimated. The angle and coordinate loss also help promote a logical relationship between measurement vectors.

\section{Experiments} \label{sec:exp}
\subsection{Datasets and Pre-processing} \label{sec:data}
LV intraventricular septum (IVS), internal diameter (LVID), and posterior wall (LVPW) dimensions were annotated in parasternal long axis 2DE scans. To avoid overfitting to a single acquisition protocol, exams were collected from four sites. All measurements were performed by a single cardiologist experienced in 2DE measurements. Diagnostic information was stripped from the images, but a mix of normal patients and varied pathologies is typical for the chosen sites. Exams were labeled at ED and ES except for where image quality in one phase prohibited accurate measurements. A total of 585 images were gathered from 309 unique patients. To generate a comparison with intra-analyser variability, 32 recordings (mixed ED and ES) were labeled multiple times by the same expert. These 64 images were set aside to be used as the test set for the network leaving 521 images for training and validation. The training, validation, and test sets were split such that images from the same patient would always remain in the same set. The coordinates and image data from the relevant frames were extracted from the stored files and converted to 256x256 one-channel images.

During training, random brightness, contrast, and gamma transformations were applied to each image. Additionally, we used mean normalization and applied random translations of 0 to 40 pixels in each direction, while ensuring coordinate locations were never within 16 pixels of the image boundaries.

\subsection{Implementation Details} \label{sec:implementation}
The network was implemented using PyTorch 0.4.1 with Python 3.6 on an Ubuntu 18.04 machine with an NVIDIA Titan X GPU. The batch size was 16 images for training and 4 images for validation. We trained for 120 epochs and reduced the learning rate by a factor of 10 every 50 epochs. Using 10\% of the training set for validation of hyperparameters, we found 4 levels was the optimal network depth and $2^6$ was the optimal number of filters in the first layer.

The primary metric important for clinical use is the accuracy of the distances for each of the three measurements. The coordinate locations of the endpoints and angle of the measurement vectors are secondary metrics that are important to create a tool that accurately follows clinical guidelines. For clinical use, it is not important that the generated heatmap matches the artificial heatmap. However, we found that keeping the relative weighting of the heatmap loss high compared to the other metrics helped improve network accuracy on all metrics. 

\subsection{Evaluation and Comparison} \label{sec:eval}

\begin{figure}[t]
  \centering
  \includegraphics[width=\textwidth]{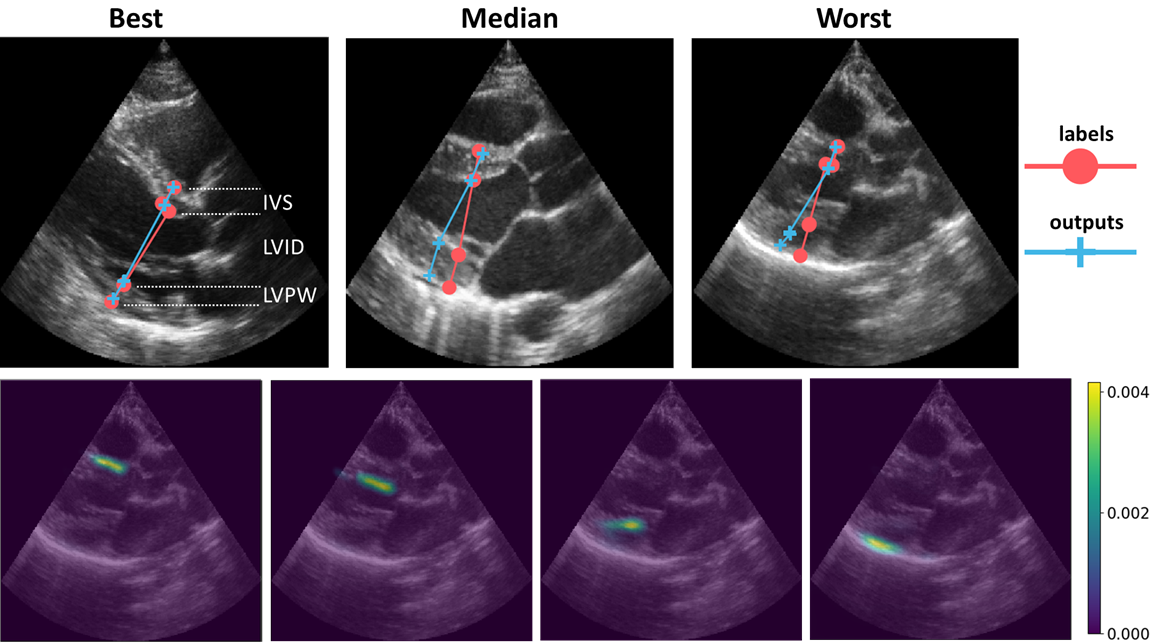}
  \caption{Top row: Qualitative results on the best, median and worst images from the test set showing expert labels and network outputs for each measurement. Bottom row: Characteristic heatmaps showing how the network learns to prioritize a small distribution in the direction parallel to the measurement direction. Only four heatmaps are shown for simplicity since the top and bottom LVID endpoints overlap with the bottom of IVS and top of LVPW respectively and produce very similar heatmaps. }
  \label{fig:results}
\end{figure}

The primary metric for evaluation was the mean percent error between the network output and ground truth distance measurements on IVS, LVID, and LVPW. The test set was composed of the 32 images that had been labeled multiple times. The median of the two labels was set as ground truth although comparing to a randomly chosen label yielded very similar results. 

While much of the strategy revolved around pre- and post- processing, we implemented several other networks in addition to U-Net for comparison. Results were compared to a stacked hourglass network \cite{B2016}, which currently obtains state of the art results on the FLIC and MPII human pose estimation metrics as well as ResNet18, ResNet34, and ResNet50 networks \cite{He2016}. We tuned the number of stacks (4) and blocks (2) of the stacked hourglass network on the validation set. We implemented the ResNet networks following the strategy proposed by Nibali et al. \cite{Nibali2018}, reducing the stride in several layers to increase output heatmap resolution, while using dilated convolutions to maintain receptive field sizes . The output heatmap size for the ResNet and stacked hourglass networks was 64x64 and we appended up-sampling layers to achieve 256x256 resolution. A CoordConv layer was added to the beginning of all networks and the same coordinate regression method and loss function were used. For a fair comparison to the other networks, results with default values of an out-of-the-box implementation of U-Net is included (no batch normalization or dropout, depth and number of filters set to 5 and $2^6$ respectively).

\section{Results} \label{sec:results}

The best, median, and worst examples (in terms of RMSE) from the test set are shown in Fig. \ref{fig:results}. The network achieves intra-analyser accuracy on LVPW and LVID measurements, and slightly worse than intra-analyser on IVS measurements. The algorithm’s worse performance on IVS measurements possibly occurs because the upper septum is often not defined as a clear gradient boundary because the septum blurs together with trabeculae in this region (see median image in Fig. \ref{fig:results}, although the network correctly found the location in this case). Expert labelers typically rely on scrolling back and forth between several frames to accurately find these points. In general, intra-analyser error is high on this task since boundaries are often blurred and lost in the noise (see the upper LVPW boundary in the worst image in Fig. \ref{fig:results} for example). The network's ability to approach intra-analyser error using only a single frame indicates that it is accurately detecting the important structures despite the high noise level. Full results on the final test set are summarized in Table \ref{tab:results}. The proposed network compares favorably to the other networks implemented on this task, achieving lower error on most metrics. We hypothesize that the performance of the other deeper networks would improve if the training dataset size were increased. However, our network has fewer parameters (which translates to a smaller memory size) and faster inference time. It is encouraging that close to expert level performance was achieved with a small network since efficient and fast implementations are important for clinical implementations.

\setlength{\tabcolsep}{12pt}
\begin{table}[]
\centering
\resizebox{\textwidth}{!}{%
\begin{tabular}{lllllll}
\toprule
\multirow{2}{*}{\textbf{Model}} & \multicolumn{4}{c}{\textbf{Mean Percent Error (\%)}}                & \multirow{2}{*}{\textbf{Params}} & \multirow{2}{*}{\textbf{Time (ms)}} \\ 
                                & \textbf{Total} & \textbf{IVS}  & \textbf{LVID} & \textbf{LVPW} &                                  &                                     \\ 
\midrule
ResNet18                        & 12.8           & 12.7          & 11.7          & 14.2          & 1e7                              & 21                                  \\
ResNet34                        & 13.0           & \textbf{11.2} & 12.1          & 15.8          & 2e7                              & 38                                  \\
ResNet50                        & 11.6           & 13.7          & 8.8           & 12.3          & 2e7                              & 43                                  \\
Stacked Hourglass               & 11.3           & 12.1          & 7.4           & 14.4          & 3e7                              & 79                                  \\
U-Net                           & 13.5           & 14.0          & 8.3           & 18.1          & 3e7                              & \textbf{10}                         \\
\textbf{Modified U-Net}         & \textbf{10.0}           & 13.4          & \textbf{6.0}  & \textbf{10.8} & \textbf{7e6}                     & 11                                  \\ 
Intra-analyser                  & 8.9            & 8.0             & 5.2             & 13.8          & n/a                              & -                                 \\ \bottomrule
\end{tabular}%
}
\caption{Comparison of proposed network to implementations of state-of-the-art networks in landmark detection and intra-analyser results. Inference time is for a single image. }
\label{tab:results}
\end{table}

\section{Conclusion} \label{sec:conclusion}
In this work we present an effective landmark detection network for 2D measurements of the LV. We demonstrate the application of these techniques in determining LV dimensions. Implementation of this network could reduce high clinical inter-/intra-analyser variability in these measurements and lead to a more repeatable diagnostic pipeline. Additionally, it enables rapid historical analysis of patients to provide robust long-term analysis. We expect that many of the techniques presented here would be applicable to other landmark detection problems in 2D and 3D ultrasound. In the future we will increase the size of the datasets, apply cross-validation, automate the detection of ED and ES frames from a full cardiac cycle, and add a confidence metric for detecting outlier results to provide a fully automated measurement tool for clinical use.

\renewcommand{\bibsection}{\section*{References}} 
\bibliographystyle{splncsnat}
\begingroup
  \ifluatex
  \else
    \microtypecontext{expansion=sloppy}
  \fi
  \small 
  \bibliography{Gilbert_2019_SUSI}

\begin{thebibliography}{12}
\providecommand{\natexlab}[1]{#1}
\providecommand{\url}[1]{\texttt{#1}}
\providecommand{\urlprefix}{}

\bibitem[{Baracho et~al.(2016)Baracho, Pinheiro, {De Melo}, and
  Coelho}]{Baracho2016}
Baracho, S., Pinheiro, D., {De Melo}, V., Coelho, R.: {A hybrid neural system
  for the automatic segmentation of the interventricular septum in
  echocardiographic images}.
\newblock Proc. Int. Jt. Conf. Neural Networks 2016-Octob, 5072--5078 (2016)

\bibitem[{He et~al.(2016)He, Zhang, Ren, and Sun}]{He2016}
He, K., Zhang, X., Ren, S., Sun, J.: {Deep Residual Learning for Image
  Recognition}.
\newblock In: CVPR. pp. 770--778 (2016)

\bibitem[{Kou and et~al.(2014)}]{Kou2014}
Kou, S., et~al.: {Echocardiographic reference ranges for normal cardiac chamber
  size: Results from the NORRE study}.
\newblock Eur. Heart J. Cardiovasc. Imaging 15(6) (2014)

\bibitem[{Lang et~al.(2015)Lang, Badano, and et~al.}]{Lang2015}
Lang, R.M., Badano, L.P., et~al.: {Recommendations for Cardiac Chamber
  Quantification by Echocardiography in Adults: An Update from the American
  Society of Echocardiography and the European Association of Cardiovascular
  Imaging}  (2015)

\bibitem[{Liu et~al.(2018)Liu, Lehman, and et~al.}]{Liu}
Liu, R., Lehman, J., et~al.: {An Intriguing Failing of Convolutional Neural
  Networks and the CoordConv Solution}  (2018)

\bibitem[{Newell et~al.(2016)Newell, Yang, and Deng}]{B2016}
Newell, A., Yang, K., Deng, J.: {Stacked Hourglass Networks for Human Pose
  Estimation}.
\newblock ECCV 9908, 483--499 (2016)

\bibitem[{Nibali et~al.(2018)Nibali, He, Morgan, and Prendergast}]{Nibali2018}
Nibali, A., He, Z., Morgan, S., Prendergast, L.: Numerical coordinate
  regression with convolutional neural networks.
\newblock CoRR abs/1801.07372 (2018)

\bibitem[{Ronneberger et~al.(2015)Ronneberger, Fischer, and Brox}]{Ronneberger}
Ronneberger, O., Fischer, P., Brox, T.: {U-Net: Convolutional Networks for
  Biomedical Image Segmentation}.
\newblock In: MICCAI (2015)

\bibitem[{Snare et~al.(2012)Snare, Mj{\o}lstad, , and et~al.}]{Snare2012}
Snare, S.R., Mj{\o}lstad, O.C., , et~al.: {Automated septum thickness
  measurement-A Kalman filter approach}.
\newblock Comput. Methods Programs Biomed. 108(2), 477--486 (2012)

\bibitem[{Sofka et~al.(2017)Sofka, Milletari, Jia, and Rothberg}]{Sofka2017}
Sofka, M., Milletari, F., Jia, J., Rothberg, A.: {Fully convolutional
  regression network for accurate detection of measurement points}.
\newblock In: DLMIA (2017)

\bibitem[{Thorstensen et~al.(2010)Thorstensen, Dalen, Amundsen, Aase, and
  Stoylen}]{Thorstensen}
Thorstensen, A., Dalen, H., Amundsen, B.H., Aase, S.A., Stoylen, A.:
  {Reproducibility in echocardiographic assessment of the left ventricular
  global and regional function, the HUNT study}.
\newblock Eur. J. Echocardiogr. 11(2), 149--156 (2010)

\bibitem[{Tompson et~al.(2015)Tompson, Goroshin, Jain, LeCun, and
  Bregler}]{Tompson}
Tompson, J., Goroshin, R., Jain, A., LeCun, Y., Bregler, C.: {Efficient object
  localization using Convolutional Networks}.
\newblock In: CVPR. vol. 07-12-June, pp. 648--656 (2015)

\end{thebibliography}
\endgroup


\end{document}